\documentclass{acm_proc_article-sp}

\begin{document}

\title{High Accuracy Gravitational Waveforms from Black Hole Binary Inspirals Using OpenCL}
%
%
%
%

\numberofauthors{3} 
%
\author{
%
%
\alignauthor
Justin McKennon\titlenote{Graduate student in the ECE Department.}\\
       \affaddr{Electrical \& Computer Engineering}\\
       \affaddr{University of Massachusetts}\\
       \affaddr{Dartmouth, MA.}\\
       \email{jmckennon@umassd.edu}
\alignauthor
Gary Forrester\titlenote{Graduate student in the Physics Department.}\\
       \affaddr{Physics Department}\\
       \affaddr{University of Massachusetts}\\
       \affaddr{Dartmouth, MA.}\\
       \email{gforrester@umassd.edu}
\alignauthor 
Gaurav Khanna\titlenote{Associate Professor in the Physics Department.}\\
       \affaddr{Physics Department}\\
       \affaddr{University of Massachusetts}\\
       \affaddr{Dartmouth, MA.}\\
       \email{gkhanna@umassd.edu}
}


\maketitle
\begin{abstract}

There is a strong need for high-accuracy and efficient modeling of extreme-mass-ratio 
binary black hole systems because these are strong sources of gravitational waves that would 
be detected by future observatories. In this article, we present sample results from our Teukolsky 
EMRI code: a time-domain Teukolsky equation solver (a linear, hyperbolic, partial differential equation 
solver using finite-differencing), that takes advantage of several mathematical and computational 
enhancements to efficiently generate long-duration and high-accuracy EMRI waveforms. 

We emphasize here the computational advances made in the context of this code. 
Currently there is considerable interest in making use of many-core processor architectures,
such as Nvidia and AMD graphics processing units (GPUs) for scientific computing. Our code uses 
the Open Computing Language (OpenCL) for taking advantage of the massive parallelism offered by 
modern GPU architectures. We present the performance of our Teukolsky EMRI code on multiple modern processor 
architectures and demonstrate the high level of accuracy and performance it is able to achieve. 
We also present the code's scaling performance on a large supercomputer i.e. NSF's XSEDE resource,  
{\em Keeneland}\footnote{A 201 TeraFLOP/s, 120-node HP SL390 system with 240 Intel Xeon 5660 CPUs 
and 360 NVIDIA Fermi M2070 graphics processors, with the nodes connected by an InfiniBand QDR network.}.  

\end{abstract}




\section{Introduction}
In the near future, a new window onto the universe will be opened. Specifically, the gravitational-wave 
spectrum that so far has been undetectable, will become observable due to the enormous investment in hardware, 
theory, and data analysis methods, such as that in the NSF LIGO\footnote{Laser Interferometer Gravitational-Wave 
Observatory: http://www.ligo.caltech.edu/} project (that is undergoing a major upgrade process) and other 
current and future detectors. 
 
This work is about the calculation of the emitted gravitational waves (GWs) from large and extreme mass-ratio 
binary inspirals (EMRIs). One of the most promising sources of low-frequency gravitational waves is the 
capture and inspiral of a compact object (such as a stellar mass black hole or a neutron star) into a 
supermasssive black hole (such as the black holes which exist at the center of many galaxies), 
following scattering processes in the core of galaxies. Such low frequency gravitational waves are expected 
to be in the good sensitivity band for space-borne gravitational wave detectors. Studies of the dynamics 
and the orbital evolution of a binary system in the extreme-mass-ratio limit are therefore an important 
issue for low-frequency gravitational wave detection.

Theoretical templates of emitted gravitational waves are necessary for detection and for parameter estimation.
Typical sources will undergo $\sim 10^5$ orbits in their last year of inspiral, and it is anticipated that 
phase coherence of the template with the signal will be required for the entire year for parameter estimation. 
This requirement drives the need for very long and highly accurate numerical simulations that are able to efficiently 
generate gravitational waveforms with relative errors lower than $10^{-4}$.  

In this article we summarize the advances made to our time-domain (TD)\footnote{As opposed to a frequency-domain 
approach that works very well in this context; except for cases where there is non-periodicity in the system, 
such as in the case of a genuine physical inspiral due to a decaying binary system.}, Teukolsky EMRI 
code~\cite{ramon,burko,pranesh,pranesh2,pranesh3,hyper} that have enabled it to achieve accuracies 
comparable to those mentioned above, while still maintaining a high degree of efficiency. 
These advances broadly lie in two distinct categories: {\em mathematical advances} and {\em computational 
performance advances}. In this article, we emphasize the latter and refer the reader to the relevant 
literature for details on the former. More specifically, we present the performance of the TD Teukolsky EMRI code 
on multiple modern processor architectures (multi-core CPUs and many-core GPUs) using the OpenCL framework and 
demonstrate the high level of accuracy and performance we are able to achieve. We also present the code's scaling  
performance on a large supercomputer with GPUs (XSEDE resource: {\em Keeneland} of Georgia Tech, Oak Ridge National Lab, 
University of Tennessee-Knoxville and the National Institute for Computational Sciences, funded by the NSF).

\section{Teukolsky EMRI Code}
Numerical Relativity (NR) is an area of computational science that emphasizes the detailed modeling of strong 
sources of GWs -- collisions of compact astrophysical objects, such as neutron stars and black holes. For the 
purposes of GW data analysis (detection and parameter estimation), it is critical to have a highly-accurate 
template bank of theoretical waveforms. Because of the degree of accuracy necessary and the large number 
of templates required, it is important to develop efficient computational methods for generating these theoretical 
waveforms. This motivates us to explore parallel computing frameworks like OpenCL and cutting-edge compute 
hardware like GPUs for NR.

The specific NR application we consider in this work is one that evolves the perturbations of a rotating (Kerr) 
black hole i.e. solves the Teukolsky equation in the time-domain. In the context of 
EMRIs, it is the small object that acts as a ``source'' of the perturbations\footnote{Because of the large 
mass-ratio, the small companion of a central supermassive black hole can be modeled as a point particle, and the problem 
can be addressed using perturbation theory.}. In other words, the Teukolsky equation 
is essentially a linear wave equation in Kerr black hole space-time geometry, with the small object acting as 
generator of the gravitational waves.

The next two subsections provide more detailed information on this equation and the associated numerical solver code.

\subsection{Teukolsky Equation}
The Teukolsky master equation describes scalar, vector and tensor field perturbations in the space-time of
Kerr black holes~\cite{eqn}. In Boyer-Lindquist coordinates, this equation takes the form
\begin{eqnarray}
\label{teuk0}
&&
-\left[\frac{(r^2 + a^2)^2 }{\Delta}-a^2\sin^2\theta\right]
         \partial_{tt}\Psi
-\frac{4 M a r}{\Delta}
         \partial_{t\phi}\Psi \nonumber \\
&&- 2s\left[r-\frac{M(r^2-a^2)}{\Delta}+ia\cos\theta\right]
         \partial_t\Psi\nonumber\\  
&&
+\,\Delta^{-s}\partial_r\left(\Delta^{s+1}\partial_r\Psi\right)
+\frac{1}{\sin\theta}\partial_\theta
\left(\sin\theta\partial_\theta\Psi\right)+\nonumber\\
&& \left[\frac{1}{\sin^2\theta}-\frac{a^2}{\Delta}\right] 
\partial_{\phi\phi}\Psi +\, 2s \left[\frac{a (r-M)}{\Delta} 
+ \frac{i \cos\theta}{\sin^2\theta}\right] \partial_\phi\Psi  \nonumber\\
&&- \left(s^2 \cot^2\theta - s \right) \Psi =  -4\pi (r^2 + a^2 \cos^2 \theta)\, T  ,
\end{eqnarray}
where $M$ is the mass of the black hole, $a$ its angular momentum per unit mass, $\Delta = r^2 - 2 M r + a^2$ and 
$s$ is the ``spin weight'' of the field. The $s = \pm 2$ versions of these equations describe the radiative degrees 
of freedom of the gravitational field, and thus are the equations of interest here. As mentioned previously, this 
equation is an example of linear, hyperbolic, partial-differential-equations (PDEs) that are quite common in several 
areas of science and engineering, and can be solved numerically using a variety of finite-difference schemes. The 
quantity $T$ in Eq. (1) is the ``source'' term as mentioned in the previous section. Ref.~\cite{eqn} has a mathematical 
formula for this quantity and to save space, we will not reproduce that expression here.

\subsection{Teukolsky Code}
To solve Eq.\ (\ref{teuk0}) numerically in time-domain we take the approach first introduced by Krivan et al. in 
Ref.~\cite{klpa}. First, we make use of Kerr spacetime's axisymmetry and factor out the $\phi$-dependence of the 
Eq.\ (\ref{teuk0}) by decomposing the solution $\Psi$ into azimuthal $m$-modes
\begin{eqnarray}
\Psi(t,r,\theta,\phi) & = & \sum_m e^{im\phi}r^3\Phi_m(t,r,\theta)\;.
\end{eqnarray} 
In this manner the Eq.\ (\ref{teuk0}) is reduced to a linear system of decoupled (2+1)-dimensional hyperbolic PDEs. 
Then, we rewrite this system in first-order form, by introducing a new auxiliary ``momentum'' field variable, $\Pi$. 
And finally, we develop an explicit time-evolution numerical scheme for this first-order, linear PDE system using the 
well-known two-step, second-order Lax-Wendroff, finite-difference method. Explicit details on this approach can be found in 
Ref.~\cite{klpa}. 

Each iteration to evolve the system above consists of two steps: In the first step, the solution vector 
$\mbox{\boldmath{$u$}}\equiv\{\Phi_R,\Phi_I,\Pi_R,\Pi_I\}$ between grid points is obtained from
\begin{eqnarray}
\label{lw1}
\mbox{\boldmath{$u$}}^{n+1/2}_{i+1/2} &=& 
\frac{1}{2} \left( \mbox{\boldmath{$u$}}^{n}_{i+1}
                  +\mbox{\boldmath{$u$}}^{n}_{i}\right)
- \\
&  &\frac{\delta t}{2}\,\left[\frac{1}{\delta r^*} \mbox{\boldmath{$D$}}^{n}_{i+1/2}
  \left(\mbox{\boldmath{$u$}}^{n}_{i+1}
                  -\mbox{\boldmath{$u$}}^{n}_{i}\right)
- \mbox{\boldmath{$S$}}^{n}_{i+1/2} \right] \; .\nonumber
\end{eqnarray}
This is used to compute the solution vector at the next time step,
\begin{equation}
\mbox{\boldmath{$u$}}^{n+1}_{i} = 
\mbox{\boldmath{$u$}}^{n}_{i}
- \delta t\, \left[\frac{1}{\delta r^*} \mbox{\boldmath{$D$}}^{n+1/2}_{i}
  \left(\mbox{\boldmath{$u$}}^{n+1/2}_{i+1/2}
                  -\mbox{\boldmath{$u$}}^{n+1/2}_{i-1/2}\right)
- \mbox{\boldmath{$S$}}^{n+1/2}_{i} \right] \, .
\label{lw2}
\end{equation}
The angular subscripts are dropped in the above equation for clarity. All angular derivatives are computed using second-order, 
centered finite difference expressions. Explicit forms for the matrices {\boldmath{$D$}} and {\boldmath{$S$}} can be easily 
found in the relevant literature~\cite{klpa}.

Symmetries of the spheroidal harmonics are used to determine the angular boundary conditions: 
For even $|m|$ modes, we have $\partial_\theta\Phi =0$ at $\theta = 0,\pi$ while $\Phi =0$ at $\theta = 0,\pi$ for 
modes of odd $|m|$. We set $\Phi$ and $\Pi$ to zero on the inner and outer radial boundaries.

\subsection{Mathematical Advances}
In this section we enlist the recent mathematical advances that have been made to the Teukolsky EMRI code that play a 
critical role in it achieving the required level of accuracy and efficiency. The reader is referred to the appropriate 
literature for details.
 
\begin{enumerate}

\item The TD Teukolsky EMRI code is a (2+1)D, linear, hyperbolic PDE solver that uses a second-order, time-explicit, 
finite-difference (two-step Lax-Wendroff) numerical evolution scheme. In the calculations, the smaller member of the 
binary is viewed as a ``source'' of  perturbations of the black hole that are evolved using the Teukolsky equation, 
that governs the evolution of perturbations in a Kerr black hole space-time. As a first approximation, one can 
construe the compact object as being pointlike and structureless. On a discrete spatial grid, a point-like source 
i.e. a {\em Dirac delta distribution} can be modeled as a smeared Gaussian distribution. Alternate and more accurate 
and efficient approaches to modeling a point-like object on a discrete grid have been developed~\cite{pranesh,pranesh2}. 
As an example, one can start by defining a step-function on a discrete numerical grid and then apply the finite-difference 
derivative operation on the discrete step-function to obtain a ``discrete-delta'' function on a computational grid. 
Results based on a refined version of this approach have shown an {\em order-of-magnitude} improvement in accuracy and 
efficiency~\cite{pranesh,pranesh2}. 

\item Another recent advance made is due to a {\em hyperboloidal} compactification for the Teukolsky equation in Kerr space-time. 
This allows one to include (null) infinity on the numerical grid by attaching a hyperboloidal layer to a compact domain 
surrounding the rotating black hole and the orbit of an inspiraling point source of GWs. This technique generates gravitational 
waveforms from large and extreme mass-ratio inspirals in Kerr space-time extracted directly at (null) infinity, while keeping the 
outer boundary location close i.e. allowing the use of a rather modest sized grid. Tests and comparisons of the results with 
previous calculations clearly establish the high-level of accuracy and efficiency of this hyperboloidal layer method as applied 
to our Teukolsky code~\cite{hyper}. 

\item Higher-order numerical methods have been used very successfully in the context of finite-difference schemes to improve 
the accuracy of the computations by further reducing the discretization error. However, it is somewhat challenging to create 
a higher-order convergent numerical-representation of a Dirac delta function and its derivatives and therefore, our code 
simply performs Richardson extrapolations of its second-order numerical solutions to obtain ones with lower discretization 
error~\cite{pranesh,pranesh2,pranesh3}. 

\end{enumerate}

\subsection{Computational Performance Advances}
Computational scientists and engineers have begun making use of many-core GPU architectures because 
these can provide significant gains in the overall performance of many numerical simulations at a relatively 
low cost. However, to the average computational scientist, these GPUs usually employ a rather unfamiliar 
and specialized programming model that often requires advanced knowledge of their architecture. In addition, 
these typically have their own vendor- and platform- specific software development frameworks (SDKs), that are 
different from the others in significant ways. For example: Nvidia's GPUs use CUDA SDK~\cite{cuda}, AMD's GPUs use 
Stream SDK~\cite{stream}, while traditional multi-core processors (from Intel, AMD, IBM) typically employ an 
OpenMP-based parallel programming model~\cite{openmp}. 

In 2009, an open standard was proposed by Apple to bring the software development for all these different 
processor architectures under a single standard -- the Open Computing Language (OpenCL)~\cite{opencl} -- and 
all major multi-core processor and GPU vendors (Nvidia, AMD, IBM, Intel) have adopted this standard for their 
current and future hardware. 

In this article, we make use of OpenCL to harness the massive parallelism offered by many-core architectures 
like GPUs in order to perform high-resolution and long-duration EMRI computations, very efficiently. This plays 
a critical role in our Teukolsky code's ability to achieve the required high level of accuracy and efficiency for EMRI 
simulations. Details are presented later in this article in Section 4.

\section{OpenCL}
As mentioned already, OpenCL is a new framework for programming across a wide variety of computer hardware architectures 
(CPU, GPU, etc). In essence, OpenCL incorporates the changes necessary to the programming language C, that allow for 
parallel computing on all these different processor architectures. In addition, it establishes numerical precision requirements 
to provide mathematical consistency across the different hardware and vendors -- a matter that is of significant importance to 
the scientific computing community. Computational scientists would need to rewrite the performance intensive routines in their 
codes as OpenCL {\em kernels} that would be executed on the compute hardware. The OpenCL API provides the programmer various 
functions from locating the OpenCL enabled hardware on a system to compiling, submitting, queuing and synchronizing the compute 
kernels on the hardware. Finally, it is the OpenCL runtime that actually executes the kernels and manages the needed data transfers 
in an efficient manner. As mentioned already, most vendors have released an OpenCL implementation for their own hardware. As of 
the writing of the document, AMD, Intel and Nvidia have OpenCL freely available for their processors. IBM has also {\em beta} 
released OpenCL for their POWER line of multi-core processors.

OpenCL is of tremendous value to the scientific community because it is open, royalty-free and 
vendor- and platform- neutral. It delivers a high degree of {\em portability} across all major forms of current 
and future compute hardware. 

\section{Code Implementation}
In this section we detail our approach taken towards parallelism, not only to take advantage of the many cores of a 
single GPU, but also those on multiple GPUs. We describe here the different ideas we have implemented and their final 
performance outcomes~\cite{opencl1,opencl2}. The lessons learned have ultimately helped us converge towards a rather 
optimal implementation.  
   
The first task in our work is to isolate the most compute intensive portions of our Teukolsky EMRI code. 
Upon performing a basic profiling of our code using the GNU profiler {\bf gprof}, we learn that computing the 
``right-hand-sides'' of the Lax-Wendroff steps i.e. the quantities within the square-brackets of Eqs.\ (\ref{lw1}) 
and (\ref{lw2}), take most of the application's overall runtime. We anticipate that this observation 
is fairly typical for codes of this type. Thus, it is natural to consider accelerating this ``right-hand-side'' 
computation using data-parallelization on the many cores of the GPU. 

A data-parallel model is relatively straightforward to implement in a code like ours. We simply perform a domain 
decomposition of our finite-difference numerical grid and allocate the different parts of the grid to different cores. 
More specifically, on the GPU, each thread computes the right-hand-side for a {\em single} pair of $r$ and $\theta$ grid 
values. In addition, it is necessary to establish the appropriate data communication between the GPU cores and the 
remaining code that is executing on the CPU -- we use {\bf  clEnqueueReadBuffer, clEnqueueWriteBuffer} instructions 
to transfer data back-and-forth from main memory and we only use {\em global memory} on the GPU to simplify communication 
between the GPU cores. We make this simplification with the goal of keeping the code's {\em portability} intact, even 
if it impacts performance to some extent~\cite{opencl2}.

Unfortunately, this naive approach yields a {\em negligible} performance gain on the GPU. The reason is that although the 
right-hand-side computation is accelerated due to the use of the many-cores of the GPU, the time it takes to bring that 
data back-and-forth from main memory so that the remaining computation can resume on the CPU, is large enough that no 
overall gain in performance is perceived. This outcome is simply due to the limited bandwidth of the system's PCI bus on which 
the GPU is located. To address this issue, we port {\em all} the Lax-Wendroff related compute routines (such as the 
computation of the evolved fields half-way between grid points, the boundary condition imposition, updating of the fields 
using the right-hand-side data) as separate kernels onto the GPU. In this manner, no communication is necessary with the 
rest of the computer system and we overcome the challenge mentioned above. It is worth noting that some of these routines 
are perhaps not ideal for execution on the GPU (for example, some don't quite have a high enough {\em arithmetic intensity} 
that is essential to obtain high performance from the GPU architecture) but we still port these over for execution on the 
GPU regardless, simply because our goal is to minimize data transfer back-and-forth from main memory. This requires a significant 
amount of additional effort -- but one that pays off well eventually (as seen in the following section). 

Porting the source-term computation i.e. the expression for $T$ onto the GPU using OpenCL was particularly challenging. The 
reason is that this expression {\em requires} complex number algebra support and because OpenCL kernels do not currently 
support C++ features, we could not simply use equivalent user-defined datatypes and operations using {\em operator overloading}. 
In operator overloading, the user specifies the operation of the mathematical operators by specifying the behavior of the operator 
on the user defined datatypes. In this light, mathematical operators are simply function calls, and the left and right hand side 
of each operator act as the arguments to these functions. Through a transformation from this standard {\em infix} notation to  
{\em prefix} notation, the operator can be forced to precede its arguments without losing the order of operations, thus forcing 
the mathematical function calls to closely mimic that of standard C-style function calls. From this point, the operators can be 
replaced by C-struct function calls by way of a recursive find and replace search tree algorithm. In this manner, complex 
arithmetic can be realized by having the C-struct functions return a struct with both real and imaginary components, preserving 
the integrity of the source-term expression.

The main limitation that we introduce with this approach of running the entire computation on the GPU is that we need 
to be able to fit the entire memory requirements of the code within the GPU video memory. Given that current high-end GPU 
offerings support only a few GBs of memory, this can be challenging. However, a compute cluster with multiple 
GPUs per node, such as NSF's XSEDE {\em Keeneland}, can overcome this serious limitation. 

\begin{figure*}
\centering
\epsfig{file=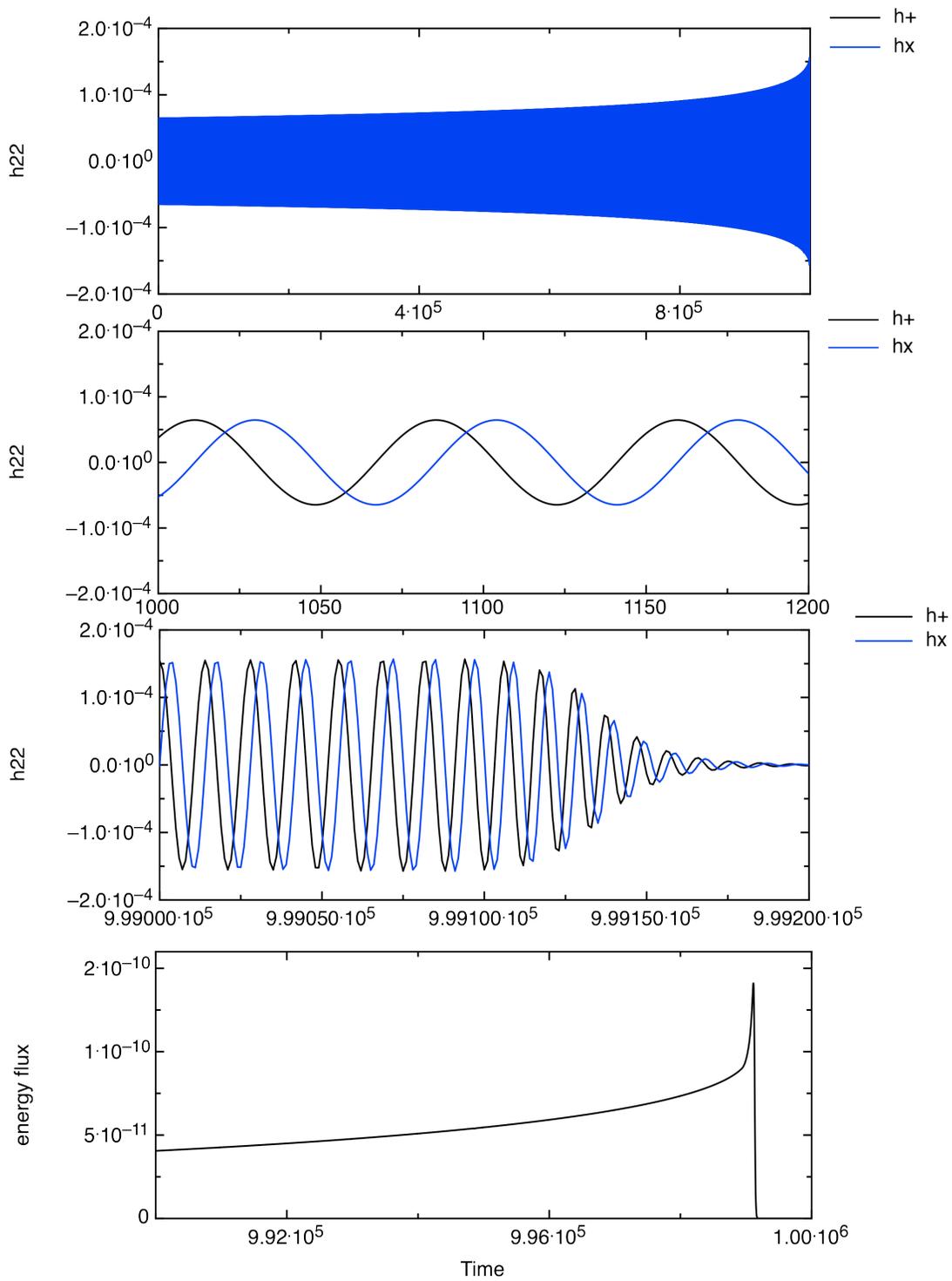,height=8in, width=6in}
\caption{{\footnotesize A complete EMRI gravitational waveform ($h_{22}$) generated using our time-domain Teukolsky EMRI code.}}
\label{full}
\end{figure*}

Another distinct approach to code parallelism that we attempted involves leaving the source-term $T$ computation on the 
multiple CPU cores and performing the rest on the many cores of the GPU. This has several advantages. First, we have full 
support for complex datatype on the CPU from standard math libraries. Second, we can make effective use of the powerful CPU 
cores during the computation, instead of leaving them mostly idle. Given the nature of the ``discrete-delta'' i.e. being 
non-zero on only a handful of grid points and the computational complexity of the source-term~\cite{opencl1}, that portion  
of the calculation is actually {\em better} suited for the few ``sophisticated'' CPU cores, as opposed to the many ``simple'' 
GPU cores. And finally, only a rather modest amount of data needs to be exchanged over the PCI bus, again, because the source-term 
is non-zero on a small number of grid points. Through detailed experimentation, we realized that this approach is quite optimal 
for our code and therefore, this is the final approach towards parallelism that we moved forward with.

It is worth mentioning that we did not make a serious attempt to hand-tune the codes to tailor them for each architecture, 
in order to obtain maximal performance. As stated earlier, one of our goals is to keep the code highly portable, because we 
aim to run the exact same code on both GPUs and CPUs. The only variable that we tuned (through simple experimentation) in order 
to obtain maximum performance for each architecture is the value of the {\em local workgroup size}. Developing the OpenCL kernels 
themselves did not require much restructuring of the original routines. Most of the effort was spent in the separating out what  
computation executes on the GPU/CPU and then setting up the communication and synchronization between them. Overall, it took 
the equivalent of two full-time (beginning) engineering graduate students working for a year to completely develop, test and 
benchmark this OpenCL code. 

Finally, to extend our parallel approach to multiple GPUs we considered the standard domain-decomposition approach using message-passing (MPI). 
However, we discovered\footnote{We thank Pranesh Sundararajan for suggesting this approach.} a novel and simpler alternative i.e. a ``temporal'' 
parallelization approach, instead of a spatial one (like domain-decomposition). 
Such a technique relies on the fact that our code is solving a linear problem, and that the trajectory for the inspiraling object is generated 
separately~\cite{pranesh2,pranesh3}. In addition, we are only interested in the ``quasi steady-state'' part of the solution (recall that code has 
a ``forcing'' source-term). Given these facts, it is possible to split the trajectory into several equal and short time-segments and then perform 
the short time-evolutions for generating the gravitational waveforms from each segment, in parallel. Then as a final step, we can ``patch'' these 
short waveforms together into a complete long waveform. There were some challenges along the way, such as handling the ``junk'' radiation burst in each 
of these short time-evolutions but there are known techniques to minimize the effect of these artifacts~\cite{jost}. This approach is particularly 
promising for {\em strong} scaling of our code on large parallel systems because it nearly eliminates all the communication necessary between 
different compute nodes. We present the scaling outcome from this approach in the next section.

\begin{figure}
\centering
\epsfig{file=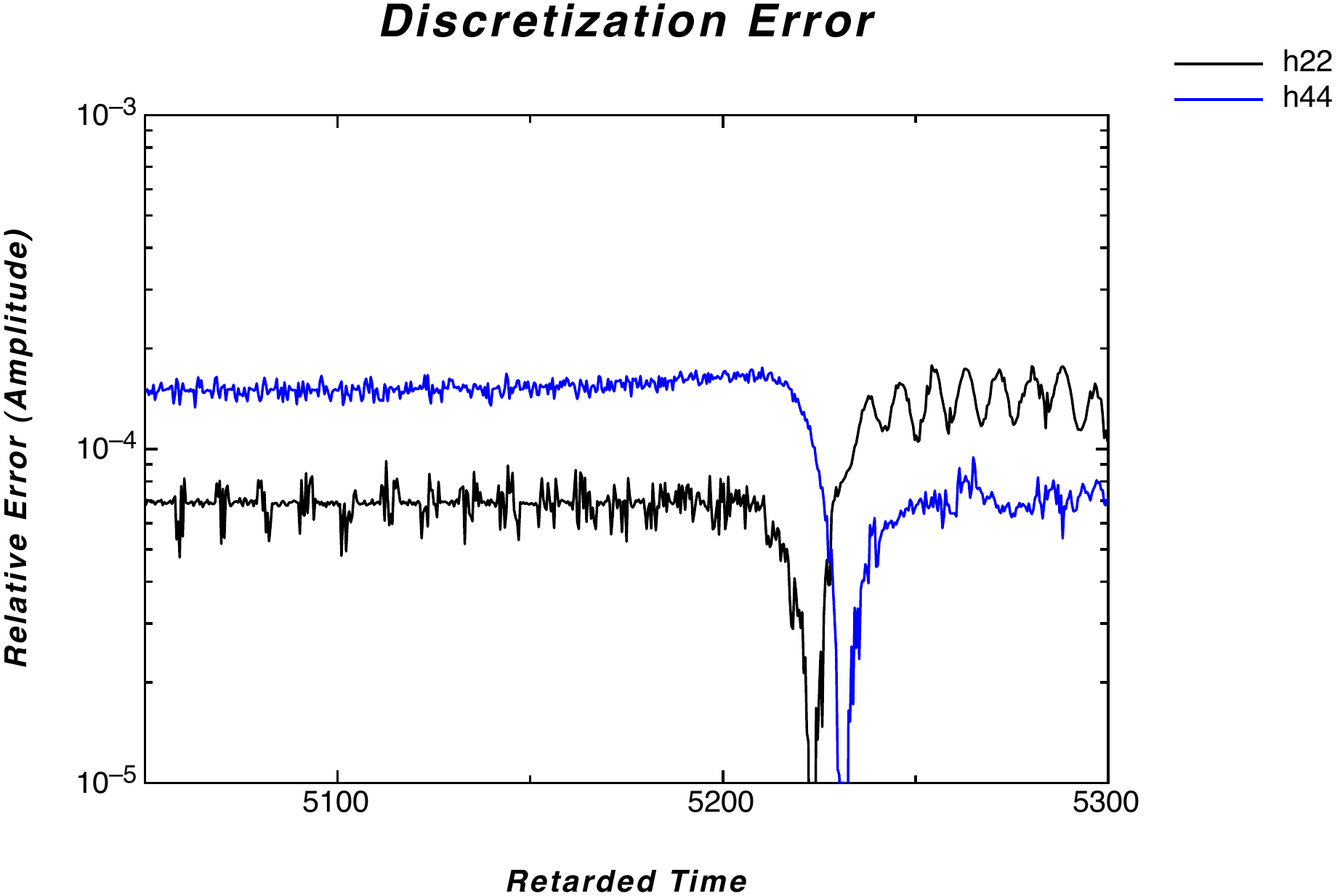,height=2.5in, width=3.5in}
\caption{{\footnotesize Relative errors in the amplitude of the $h_{22}$ and $h_{44}$ modes of the gravitational waveform.
Compare with Figure 1 in Ref.~\cite{myeob}.}}
\label{amp}
\end{figure}

\section{Results}
This section is dedicated to the outcomes from the mathematical and computational advances made to our Teukolsky 
EMRI code detailed in the previous sections. 

Figure \ref{full} presents a complete EMRI gravitational waveform generated using our time-domain Teukolsky EMRI code. 
The mass-ratio used for this evolution is $10^{-4}$ and the (circular, equatorial) orbital decay covers all phases, from the 
adiabatic inspiral to the plunge regime. The duration of the inspiral is a {\em million} M long i.e. over 10,000 full 
orbital cycles. The data shown in the plot is the $\ell = m = 2$ ``spin-weighted'' spherical harmonic projection of the 
full gravitational wave strain i.e. $h_{22}$.  

\subsection{Accuracy}
\begin{figure}
\centering
\epsfig{file=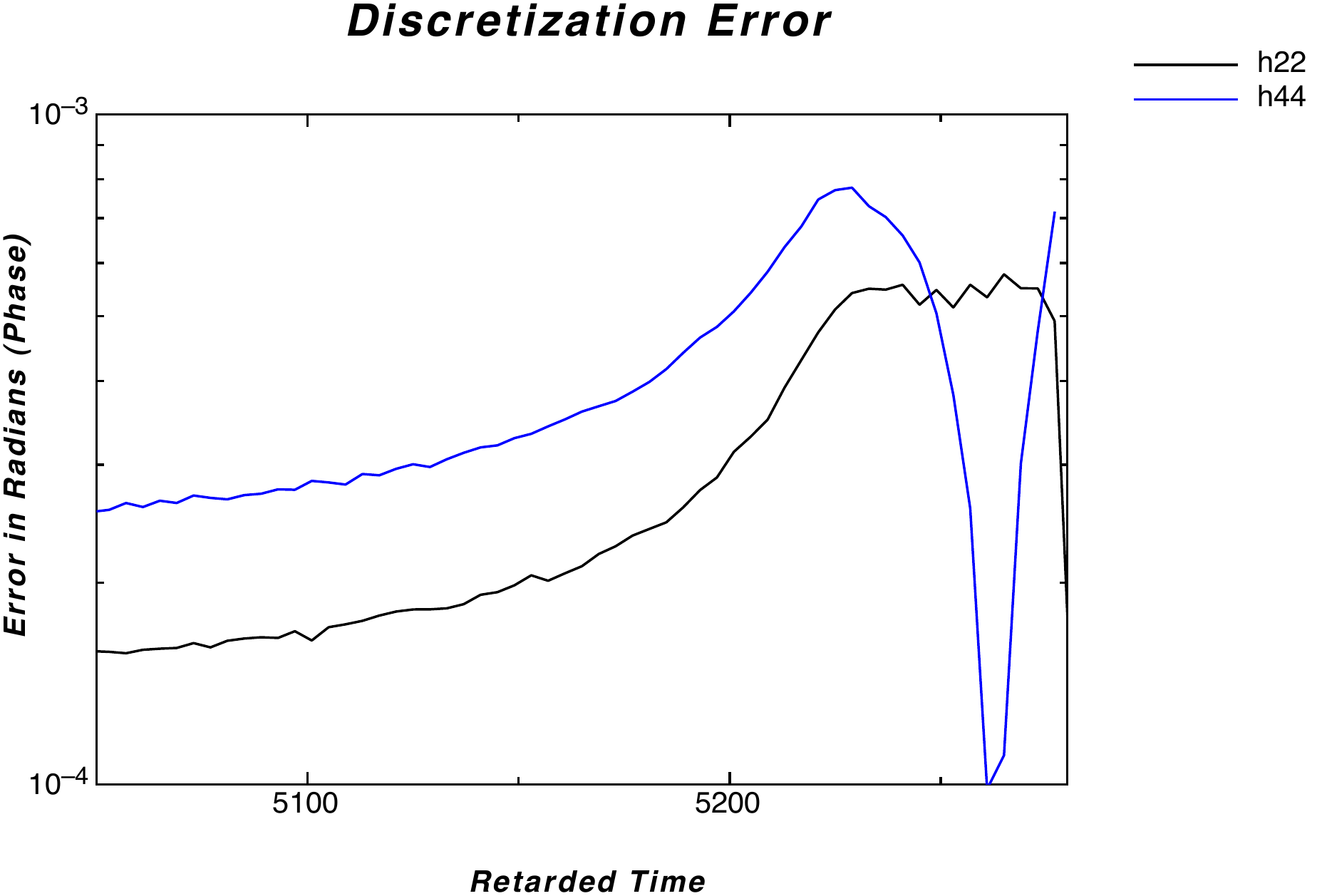,height=2.5in, width=3.5in}
\caption{{\footnotesize Error in the phase of the $h_{22}$ and $h_{44}$ modes of the gravitational waveform.
Compare with Figure 1 in Ref.~\cite{myeob}.}}
\label{phase}
\end{figure}

In Figures \ref{amp} and \ref{phase} we depict the discretization errors from a sample high-accuracy computation\footnote{
This error estimate is obtained by computing the waveform data's Richardson extrapolant two different ways and then 
taking the relative difference between these two. The highest $(r,\theta)$ resolution in use here is $(M/320,\pi/200)$.}. 
It is clear that for the most part the errors stay at acceptably low levels (on the scale of $10^{-4}$). However, towards 
the tail end of the computation, the errors do grow to somewhat higher levels. This happens due to a dramatic change in 
the nature of the computation at late times. More specifically, the inspiralling compact object plunges into the central 
black hole and thus rapidly ``disappears'' from the computational domain, thereby transitioning the Teukolsky equation 
from one that is strongly source-term dominated, into its homogeneous form. This effect causes a modest change in the 
convergence rate of the code which ultimately reflects in the error plots depicted here.  

{\em It is worth noting that these error levels are over an order-of-magnitude lower than those reported before}~\cite{myeob}. 
The combination of our mathematical and computational advances have made this high-level of accuracy feasible. 

\subsection{Code Performance: Single GPU}
Table \ref{comp} depicts the relative values for overall performance of our Teukolsky EMRI code for several variants of current 
generation CPUs and GPUs. These results suggest that it is relatively straightforward to obtain {\em order-of-magnitude} gains 
in overall code performance by making use of many-core GPUs over multi-core CPUs and this fact is largely independent of the 
specific hardware architecture and vendor. All the systems used in these performance tests used a variant of the Linux operating 
system and OpenCL provided by the appropriate vendor (AMD\footnote{Catalyst 12.4; APP SDK 2.6} or Nvidia\footnote{CUDA 4.0}). 
Detailed specifications of the compute hardware are included in the table. The $(r,\theta)$ grid size for this performance study  
is $32000 \times 304$ that nearly utilizes the entire global memory capacity (3 GBs) of the considered GPUs. We use full 
{\em double-precision} floating point accuracy in all our computations. The emphasis on high-accuracy in this work requires us 
to make use of the highest grid resolution and numerical precision offered by the compute hardware.

\begin{table}
  \centering
     \begin{tabular}{|c|c|c|c|c|} \hline
   {\bf Name} & {\bf Type} & {\bf GHz} & {\bf Cores} & {\bf Perf.}\cr
    \hline \hline
    {\bf AMD Opteron 6200}   & CPU &   2.1    &  16    & {\bf 1x} \cr \hline
    {\bf Intel Xeon E5-2600} & CPU &   2.2    &  16    & {\bf 2.2x} \cr \hline
    {\bf Nvidia Fermi 2050}  & GPU &   1.2    &  448   & {\bf 13x} \cr \hline
    {\bf AMD Radeon 7970}    & GPU &   1.0    &  2000  & {\bf 19x} \cr \hline
   \end{tabular}
 \caption{{\footnotesize This table depicts the relative values for overall performance for several variants of current generation 
CPUs and GPUs. The baseline system here has dual AMD Opteron, 8-core, 2.1 GHz CPUs running our OpenCL Teukolsky code. The remaining 
hardware listed above is co-located in a separate system.}}
\label{comp}
\end{table}

It is also noteworthy that the consumer-grade GPU, the AMD Radeon HD 7970, outperforms Nvidia's HPC-oriented, high-end  
Fermi M2050 GPU, while maintaining a significantly lower cost\footnote{We are aware that with the release of the M2090 GPU, the 
M2050 is no longer the highest-grade Nvidia Fermi GPU anymore. However, we estimate that the Radeon 7970 would still outperform 
the M2090 on our tests.}. {\em The cost effectiveness of such consumer-grade compute 
hardware is nearly an order-of-magnitude higher than the alternatives}. This observation is consistent with our earlier work that 
evaluated the {\em Sony PlayStation 3} consumer gaming console for scientific computing~\cite{opencl1,ps3}.

\subsection{Code Performance: Multiple GPUs}
\begin{figure}
\centering
\epsfig{file=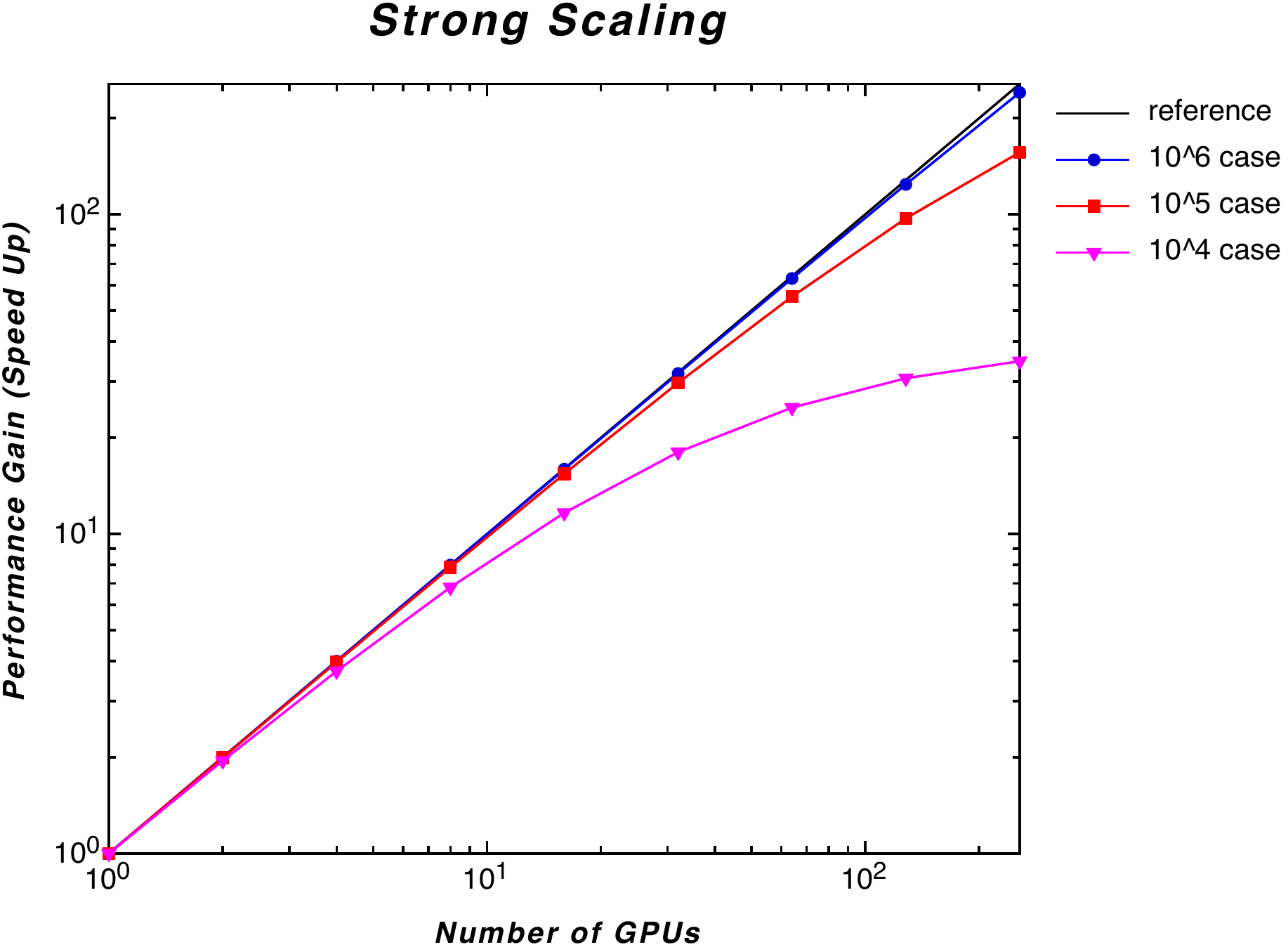,height=2.5in, width=3.5in}
\caption{{\footnotesize Strong scaling on the NSF XSEDE Keeneland supercomputer for several long-duration simulations.}}
\label{scaling}
\end{figure}

Figure \ref{scaling} depicts the outcome of a strong scaling study performed on the XSEDE {\em Keeneland} system 
using our Teukolsky EMRI code. The message-passing based parallelization approach 
we take is explained in Section 4 of this article. The longer-duration EMRI computations scale much better, in particular {\em the $10^{6}M$ 
long simulation scales almost perfectly}. This is simply because, our parallel approach requires us to overlap the computations 
running on each GPU by a fixed amount, in order to perform the suitable ``patching'' during the post-processing stage. And 
because this overlap is of fixed duration, we start to receive diminished returns, especially for the cases with the larger 
number of GPUs. Of course, this overlapping is negligible for the very long duration cases, which is why the scaling is much 
better for those. 

\subsection{Science Enabled}
In this section we document the current significant scientific contributions that have been made to the 
general area of gravitational physics (from gravitation wave physics to issues closely associated to 
Cosmic Censorship and also black hole astrophysics) owing to the development of our efficient and 
high-accuracy time-domain Teukolsky EMRI code. 

\begin{enumerate}

\item {\em Waveform generation}: As pointed out in the previous sections, the advancements made to the TD Teukolsky EMRI code enable the code to perform very high accuracy and 
very long duration evolutions in a very efficient manner. As an example, in Ref.~\cite{hyper} we demonstrate relative errors in the gravitational wave energy flux at null 
infinity on the scale of $10^{-4}$ and also perform a {\em million} $M$ long EMRI evolution (over $10,000$ orbital cycles -- see Fig.~\ref{full}).

\item {\em Calibration of EOB models}: Effective-one-body~\cite{eob} formalism is an analytical approach that can very efficiently model black hole binary systems over a wide 
range of mass-ratios (from comparable to extreme) and is thus perhaps best suited for the generation of the large banks of templates needed for gravitational wave data 
analysis. High accuracy results from our code have contributed towards the development of a ``calibrated'' EOB model for large mass-ratio  binaries with spin~\cite{myeob} in 
the context of quasi-circular, equatorial orbits. 

\item {\em Recoil ``kick'' velocities}: Gravitational waves carry away linear momentum from a decaying binary, thus causing the system to recoil or ``kick''~\cite{kick}. 
A peculiar aspect of the recoil is that in certain scenarios (a prograde orbit decaying around a rapidly rotating black hole, in the context of large mass-ratios) there is a 
large ``anti-kick'' that appears very late in the plunge phase, that seems to completely cancel the large accumulated recoil present up to that point~\cite{pranesh3}. While 
there have been several mechanisms that have been proposed for this phenomenon, they all involve a significant role played by horizon perturbations i.e. quasi-normal ringing 
of the black hole. Recent work made possible due to the TD Teukolsky EMRI code suggests otherwise, and proposes a much simpler mechanism for the origin of the anti-kick~\cite{mykick}. 
In addition, as a by-product, the work also developed a scheme (``integration-from-peak'') using which one can obtain excellent estimates for kicks from very short evolutions, 
thus potentially being of great value to full NR. 

\item {\em Cosmic Censorship}: It has been suggested multiple times in literature (see Ref.~\cite{jacobson} for a recent proposal) that it may be possible for a near-extremal 
Kerr black hole to capture a test particle (on a specifically designed trajectory) that would result in overspinning the hole, thus forming a naked singularity in violation of 
the Cosmic Censorship Conjecture. The general expectation has been that once one accounts for radiation-reaction, there would be no overspinning and therefore Cosmic 
Censorship would be preserved. We have now been able to demonstrate, through the use of the TD Teukolsky EMRI code, that it is actually the effect of the {\em conservative} 
part of the gravitational self-force that is likely responsible for preventing the overspinning~\cite{naked,naked2}. One way to describe the outcome of this research is that a 
near extremal Kerr hole simply fails to capture a test particle that could potentially overspin it!  

\end{enumerate}

\section{Conclusions}
In this article we demonstrate that recent mathematical and computational advances made to our time-domain Teukolsky EMRI code have enabled it to achieve a high-level 
of accuracy and efficiency. We emphasize the computational advancements made, that make use of the OpenCL framework to take advantage of the massive parallelism offered 
by modern many-core GPU architectures. The {\em order-of-magnitude} gain in computational performance we obtain in this manner plays a critical role in our code achieving 
the desired level of accuracy and efficiency.  

The ability to perform high-accuracy and long-duration EMRI computations has enabled various interesting advances in gravitational physics. Using data generated by 
this code we have been able to make significant contributions to the development of effective-one-body models and gravitational waveform generation, that will ultimately 
positively impact the data analysis of current and future detectors (such as NSF's LIGO and future space-borne missions). In addition, results from our code have brought 
forth a better understanding of the ``anti-kick'' which is an intriguing aspect of the phenomenon of gravitational recoil in decaying binary systems due to gravitational 
wave emission. And finally, our code has also helped test Cosmic Censorship in the context of the capture of a small test particle by a near extremal Kerr black hole.

\section{Acknowledgments}
We would like to thank Mike DeSousa for double-checking several the accuracy level tests that we performed with the code. We are grateful to Glenn Volkema, Steve Liebling and Mark Barnell for providing useful feedback on an earlier version of this manuscript, and to the HPC group at MICROWAY for providing us access to some of the hardware we used in our study. We acknowledge research support from NSF Grant Nos. PHY-0902026, CNS-0959382, PHY-1016906 and PHY-1135664, and AFOSR DURIP Grant No. FA9550-10-1-0354 and also the Massachusetts Space Grant Consortium. Most of the numerical simulations needed for this work were performed on the NSF XSEDE {\em Keeneland} supercomputer under project number UT-NTNL0036.


\begin{thebibliography}{1}

\bibitem{ramon} R. Lopez-Aleman, G. Khanna and J. Pullin: ``Perturbative evolution of particle orbits around Kerr black holes: time domain calculation'', Class. Quantum Grav. {\bf 20}, 3259 (2003).


\bibitem{burko} L. M. Burko and G. Khanna: ``Accurate time-domain gravitational waveforms for extreme-mass-ratio binaries'',  Europhys. Lett. {\bf 78}, 60005 (2007). 

\bibitem{pranesh}  P. A. Sundararajan, G. Khanna, and S. A. Hughes: ``Towards adiabatic waveforms for inspiral into Kerr black holes: I. A new model of the source for the time domain perturbation equation'', Phys. Rev. D {\bf 76}, 104005 (2007).

\bibitem{pranesh2}  P. A. Sundararajan, G. Khanna, S. A. Hughes, S. Drasco: ``Towards adiabatic waveforms for inspiral into Kerr black holes: II. Dynamical sources and generic orbits'', Phys. Rev. D {\bf 78}, 024022 (2008).

\bibitem{pranesh3}  P. A. Sundararajan, G. Khanna, S. A. Hughes: ``Binary black hole merger gravitational waves and recoil in the large mass ratio limit'', Phys. Rev. D {\bf 81}, 104009 (2010).

\bibitem{hyper} A. Zenginoglu, G. Khanna: ``Null infinity waveforms from extreme-mass-ratio inspirals in Kerr spacetime'', Phys. Rev. X {\bf 1}, 021017 (2011).

\bibitem{eqn} 
S. Teukolsky: ``Perturbations of a rotating black hole'', Astrophys. J. {\bf 185} 635 (1973).

\bibitem{klpa} 
W. Krivan, P. Laguna, P. Papadopoulos, and N. Andersson: ``Dynamics of perturbations of rotating black holes'', Phys. Rev. D {\bf 56}, 3395 (1997). 

\bibitem{cuda}
Nvidia's CUDA \textsf{http://www.nvidia.com/cuda/}

\bibitem{stream}
AMD's Stream \textsf{http://www.amd.com/stream}

\bibitem{openmp}
OpenMP \textsf{http://openmp.org}

\bibitem{opencl} 
Khronos OpenCL \textsf{http://www.khronos.org/opencl/}

\bibitem{opencl1} G. Khanna, J. McKennon: ``Numerical modeling of gravitational wave sources accelerated by OpenCL'', Computer Physics Communications {\bf 181}, 1549 (2010).

\bibitem{opencl2} N. Choudhary, R. Ginjupalli, S. Navada and G. Khanna: ``An Exploration of OpenCL for a Numerical Relativity Application'', Proceedings of the Parallel and Distributed Computing Systems (PDCS) conference, Dallas, TX (2011).

\bibitem{eob}  A. Buonanno and T. Damour: ``Effective one-body approach to general relativistic two-body dynamics'', Phys. Rev. D {\bf 59}, 084006 (1999).

\bibitem{myeob} E. Barausse, A. Buonanno, S. A. Hughes, G. Khanna, S. O'Sullivan and Y. Pan: ``Modeling multipolar gravitational-wave emission from small mass-ratio mergers'',  Phys. Rev. D {\bf 85}, 024046 (2012).

\bibitem{kick} M. Favata, S. A. Hughes, D. Holz: ``How black holes get their kicks: Gravitational radiation recoil revisited'', Astrophys. J. {\bf 607}, L5 (2004).

\bibitem{mykick} R. H. Price, G. Khanna and S. A. Hughes: ``Systematics of black hole binary inspiral kicks and the slowness approximation'',  Phys. Rev. D {\bf 83}, 124002 (2011).

\bibitem{jacobson} T. Jacobson and T. Sotiriou: ``Overspinning a Black Hole with a Test Body'', Phys. Rev. Lett. {\bf 103}, 141101 (2009).

\bibitem{naked} E. Barausse, V. Cardoso and G. Khanna: ``Test bodies and naked singularities: is the self-force the cosmic censor?'', Phys. Rev. Lett. {\bf 105}, 261102 (2010).

\bibitem{naked2} E. Barausse, V. Cardoso and G. Khanna: ``Testing the Cosmic Censorship Conjecture with point particles: the effect of radiation reaction and the self-force'', Phys. Rev. D {\bf 84}, 104006 (2011).

\bibitem{jost} S. Field, J. Hesthaven, S. Lau: ``Persistent junk solutions in time-domain modeling of extreme mass ratio binaries'', {\em preprint} arXiv:1001.2578 (2010).

\bibitem{ps3} PS3 Gravity Grid \textsf{http://gravity.phy.umassd.edu/ps3.html}

\end{thebibliography}
\end{document}